\documentclass{ws-m3as}
\usepackage{amssymb, amsmath, amsthm,dsfont}
\usepackage{color}
\usepackage{mathtools}
\usepackage{nicefrac}
\usepackage{csquotes}
\usepackage{amsmath,amssymb,amsthm,mathrsfs,amsfonts}
\newtheorem{thm}{Theorem}[section]

\newtheorem{lem}[thm]{Lemma}

\theoremstyle{remark}
\newtheorem{rem}[thm]{Remark}

\theoremstyle{definition}
\newtheorem{defn}{Definition}[section]

\usepackage{multicol}
\usepackage{blindtext}

\usepackage{subfig}
\usepackage[super]{natbib}
\setcitestyle{citesep={,}}

\begin{document}

\markboth{G. Estrada-Rodriguez, H. Gimperlein, K. J. Painter \& J. Stocek}{Space-time fractional diffusion in cell movement with delay}

\catchline{}{}{}{}{}

\title{SPACE-TIME FRACTIONAL DIFFUSION IN CELL MOVEMENT MODELS WITH DELAY}  

\author{GISSELL ESTRADA-RODRIGUEZ}

\address{Maxwell Institute for Mathematical Sciences and Department of Mathematics, Heriot--Watt University\\
Edinburgh, EH14 4AS, United Kingdom\\
ge5@hw.ac.uk.}

\author{HEIKO GIMPERLEIN}

\address{Institute for Mathematics, University of Paderborn\\
Warburger Str.~100, 33098 Paderborn, Germany\\
Maxwell Institute for Mathematical Sciences and Department of Mathematics, Heriot--Watt University\\
Edinburgh, EH14 4AS, United Kingdom\\
h.gimperlein@hw.ac.uk.}

\author{KEVIN J. PAINTER}

\address{Maxwell Institute for Mathematical Sciences and Department of Mathematics, Heriot--Watt University\\
Edinburgh, EH14 4AS, United Kingdom\\
k.painter@hw.ac.uk.}

\author{JAKUB STOCEK}

\address{Maxwell Institute for Mathematical Sciences and Department of Mathematics, Heriot--Watt University\\
Edinburgh, EH14 4AS, United Kingdom\\
js325@hw.ac.uk.}
 
\maketitle 

\begin{history}
\received{(Day Month Year)}
\revised{(Day Month Year)}
\comby{(xxxxxxxxxx)}
\end{history}

\begin{abstract}
\noindent The movement of organisms and cells can be governed by occasional long distance runs, 
according to an approximate L\'{e}vy walk. For T cells migrating through chronically-infected brain 
tissue, runs are further interrupted by long pauses and the aim here is to clarify 
the form of continuous model equations that describe such movements. Starting from a microscopic
velocity-jump model based on experimental observations, we include power-law distributions of run and waiting times and investigate the relevant parabolic limit from a kinetic equation for resting 
and moving individuals. In biologically relevant regimes we derive nonlocal diffusion equations, 
including fractional Laplacians in space and fractional time derivatives. Its analysis and 
numerical experiments shed light on how the searching strategy, and the impact from chemokinesis
responses to chemokines, shorten the average time taken to find rare targets in the absence of 
direct guidance information such as chemotaxis.   
\end{abstract}



\vskip 1.0cm

\section{Introduction}\label{intro}

Modelling biological movement has received significant attention, with a large body of 
work devoted to deriving macroscopic (PDE) equations for the mean 
behaviour of some underlying microscopic movement model. A common description 
assumes movement follows a velocity-jump random walk, an alternating sequence of runs (movement 
with a fixed velocity) and reorientations (choosing a new velocity). When the movement 
is subject to an external bias, such as a chemical attractant, a series of studies 
dating to Patlak \cite{patlak1953random} has generated solid understanding on how 
microscopic detail translates into a diffusion-advection 
type equation. \cite{bellomo2008modeling, othmer2013} Many derivations follow a fairly standard 
set of assumptions on individual behaviour, such as negligible waiting times between jumps 
and that the distribution of runtimes follows a Poisson distribution, as observed for classic 
studies on cells such as {\em E. coli}. \cite{berg1972chemotaxis}
Under these assumptions, the macroscopic diffusion is of classic Fickian form. 

Yet these assumptions do not apply universally, such as when 
searching for sparsely distributed targets. Recent years have witnessed
reports on the tendency towards long-range diffusion, where a particle's motion 
follows the characteristics of a L{\'{e}}vy flight: occasional non-localised flights 
that interrupt local movements. Intuitively, the probability of remaining stuck in 
non-productive regions decreases and the mean time taken to find rare targets is reduced. 
Non-Brownian search strategies have been reported for microorganisms, including 
\textit{E.~coli} \cite{korobkova2004molecular} and {\em Dictyostelium}, \cite{li2008persistent} 
immune cells, \cite{harris} and large organisms (e.g. mussels, \cite{de2011levy} marine predators \cite{humphries2010environmental,sims2008scaling} and monkeys \cite{ramos2004levy}). The natural strategies have been adopted for robots. \cite{KrivonosovDZ16}

Motivated by the movements of immune cells in chronically infected brain tissue, \cite{harris} 
here we derive the macroscopic model for a microscopic velocity-jump random walk 
(Section \ref{sec: micro}) in which both the runtime distance and waiting time between 
re-orientations follow long-tail (approximate L\'{e}vy) distributions. 
The delay is the key new ingredient from a modelling perspective, observed in experiments. 
\cite{harris, miller} We derive the appropriate kinetic-transport equation, where the 
\enquote{collision} term describes the nonlocal motion. Solving an equation for the 
resting population introduces a nonlocal delay in time for the moving population and, via 
a perturbation argument and appropriate space/time scaling, obtain the following 
nonlocal equation for the population density ($u_\textnormal{tot}$):
\begin{equation}
{}_{t}^C\mathds{D}^\kappa  u_\textnormal{tot}=\nabla\cdot\left(C_{\alpha,\kappa}\nabla^{\alpha-1}u_\textnormal{tot} \right)\label{eq: final in intro}\ .
\end{equation}
In the above, ${}_{t}^C\mathds{D}^\kappa$ is the fractional time derivative in the sense of Caputo, 
$\kappa \in (0,1)$, while $\nabla^{\alpha-1}$ denotes a fractional gradient for $\alpha\in(1,2)$, \textcolor{black}{see the Definition in \ref{def: fractional}}. In the physical regime 
$\frac{\alpha}{\kappa} \in [1,2]$: this ranges from ballistic motion for $\alpha= \kappa=1$, with 
a resulting fractional heat equation governed by a L\'{e}vy process, to standard diffusion 
 for $\alpha=2$, $\kappa=1$. The population is governed by a diffusion term with 
coefficient $C_{\alpha,\kappa}$ (defined at the end of Section \ref{section:finalequation}) that 
represents a random component to motility. As described in greater detail below, 
experimental data on immune cell movements lead to $\alpha = 1.15$ and $\kappa = 0.7$. \cite{harris} 
While our approach is often applied in the context of chemotaxis, it is noted that 
\eqref{eq: final in intro} does not contain a chemotactic component; this is in agreement 
with Ref.~\refcite{harris} where the 
immune cells do not appear to exhibit directional migration on the experimental time/length scales. 

The simple structure of Eq.~\eqref{eq: final in intro} allows analytic insights 
not directly visible from the microscopic model. In particular, in Section 
\ref{sec: fundamental solution} we explicitly write down the fundamental solution 
in $\mathbb{R}^d$ and, as direct applications, we discuss hitting and mean first passage 
times. Numerical experiments are presented in Section \ref{numerics}, allowing 
efficient quantitative description and a basis for parametric studies into immune cell 
search strategies.

\section{Background and data}\label{bio}

{\em Toxoplasma gondii} ({\em T. gondii}) is a species-crossing parasitic 
pathogen \cite{blanchard2015persistence} with high seroprevalence in humans. Acute infection
is followed by chronic infection, with the parasite taking up 
lifetime residence in the host’s central nervous system (CNS). While regarded 
as generally symptomless, infected individuals with compromised immune systems are at 
greater risk of life-threatening recurrence and chronic infection has also been 
linked to altered neurological behaviour. \cite{parlog2015toxoplasma} Long term 
immunity and control of chronic {\em T. gondii} infection primarily relies on 
CD8$^+$  T cells, \cite{hwang2015cd8} which continuously search for and eliminate 
infected cells through contact. A recent study of CD8$^+$  dynamics in infected brain 
tissue has revealed a number of insights into their chemical control and movement 
patterns. \cite{harris} At a chemical level, the CXCL10/CXCR3 chemokine signalling 
system controls both the initial recruitment and subsequent maintenance of a 
CD8$^+$  population \cite{harris}: anti-CXCL10 treatments lower the resident 
population of T cells and increase parasite densities. Further, CXCL10 appears to act 
as a chemokinetic agent during the chronic phase, with anti-CXCL10 treatment reducing 
average cell velocities. \cite{harris}

\begin{figure}[ht]
            \centering
         \subfloat[\label{fig: Levy walk2}]{\includegraphics[width = 0.52\textwidth]{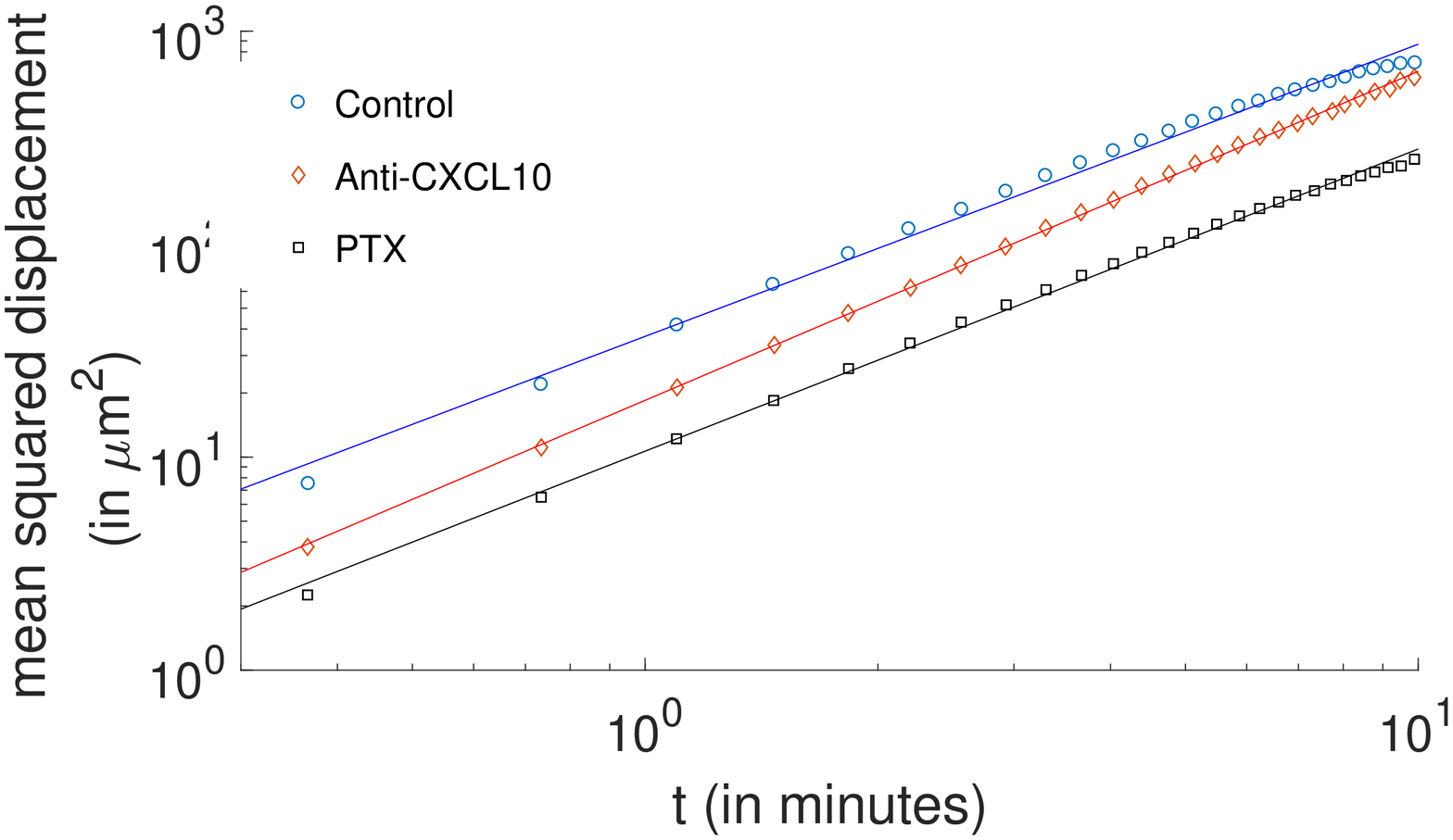}}
        \subfloat[\label{fig: Mittag-Levy2}]{\includegraphics[width=0.52\textwidth]{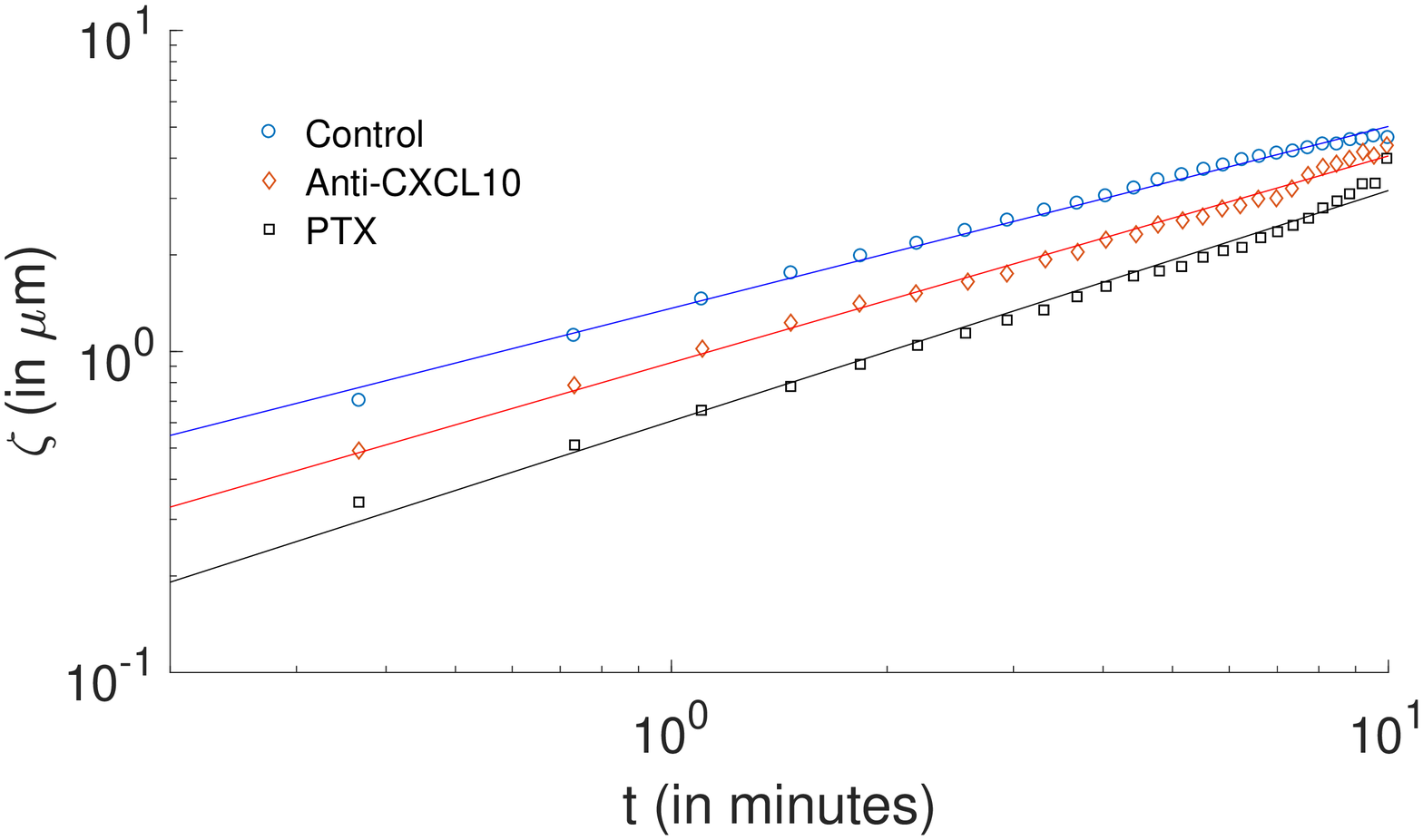}}
        \caption{Reproduction of CD8$^+$  T cell tracking data in \cite{harris}, indicating generalized 
        L\'{e}vy diffusive behaviour in the central nervous system. (a) mean squared distance for $CD8^+$ T
        cells in control tissue (blue) and two treatments that impact on chemokine signalling (mice 
        treated with anti-CXCL10 antibodies, red, and mice treated with PTX, a chemokine 
        signalling inhibitor, black). (b) Spatial scaling factor of the self-similar diffusion.}
\end{figure}\label{fig: comparison2}    

Analyses of CD8$^+$ T cell tracks in Ref.~\refcite{harris} suggests that they 
follow a generalised L{\'{e}}vy walk. We reproduce the mean squared distances showing 
superdiffusive behaviour ($\langle x^2\rangle \sim t^{1.4}$) in Figure \ref{fig: Levy walk2}. 
Yet, dependence of the spatial scaling on time (Figure \ref{fig: Mittag-Levy2}) is inconsistent 
with a L\'{e}vy walk in the absence of waiting times. In Ref.~\refcite{harris} various models for 
T cell migration are examined, including random walks, persistent random walks and 
L\'{e}vy walks, with the conclusion that the experimental results are best described by 
a generalized L\'{e}vy walk. The microscopic description is as follows: (1) cells make 
straight runs with fixed velocity but random orientation, where the run distance 
is chosen randomly from a L\'{e}vy distribution ($L_\mu(\ell)\sim\ell^{-\mu}$) with 
exponent $\mu_{run}=2.15$; (2) following each run, cells pause for a time that is also 
distributed according to a L\'{e}vy distribution with exponent $\mu_{pause}=1.7$. 
L\'{e}vy distributions for the distance $\ell$ and times $\tau$ are drawn from the 
following expressions
\[
Z_{\mu}=\frac{\sin((\mu-1)X)}{(\cos X)^{1/(\mu-1)}}\left(\frac{\cos((2-\mu)X)}{Y} \right)^{(2-\mu)/(\mu-1)}
\]
where $X$ is a uniform random variable on the interval $[-\pi/2,\pi/2]$ and $Y=-\ln X'$. 
For runs, once a distance $\ell$ is chosen, the walker moves in a randomly chosen 
direction for a time $\ell/v$, where $v$ is the velocity of the walker. For pauses, 
once a time $\tau$ is chosen, the walker remains stationary for that length of time.

While anti-CXCL10 treatments reduce CD8$^+$ T cell speed and/or increase pauses, other 
migration statistics of the T cells remain the same: $\mu_{run}=2.15$ and $\mu_{pause}=1.7$, 
as in the control case. Thus, CXCL10 appears to operate as a chemokinetic agent through 
increasing the rate at which patrolling CD8$^+$ T cells encounter their sparsely distributed targets, 
with CXCL10 (and other chemokines) shortening capture time through 
faster movement speeds. \cite{harris}

\section{Microscopic model description}\label{sec: micro}

We model a population of CD8$^{+}$ T cells moving in a medium in $\mathbb{R}^n$. It is noted
that for the experimental system of Ref.~\refcite{harris}, the resident T cell population 
numbers somewhere between $300,000$ and $450,000$ across a volume of 
$3.2-4.4\times 10^{11}\mu m^3$, motivating a continuum description for their 
collective movement. Microscopically, we assume each individual performs a 
generalized L\'{e}vy walk with the following properties:
\begin{enumerate}
\item The interactions between individuals are taken to be negligible. This assumption appears 
reasonable, given the relatively low densities of T cells.
\item Starting at position $\mathbf{x}$ and time $t$, we assume an individual runs in 
direction $\theta$ for some time $\tau$, called the \enquote{run time}. This run time is selected
from a distribution $\psi$.
\item During runs, individuals are assumed to move with constant forward speed $c$ and take a straight 
line motion between reorientations. 
\item Each time the individual stops it selects a new direction $\eta$ according to 
a distribution $k(\mathbf{x},t,\mathbf{\theta};\mathbf{\eta})$ which only depends on $|\theta - \eta|$, after waiting for some time $r$. The choice of new direction is taken to be independent 
of chemical concentrations/gradients. 
\item The reorientation time $r$ follows a L\'{e}vy distribution $\psi_r(r)$. 
\end{enumerate}
Note that assumptions (3-4) derive from the experimental conditions of CD8$^+$ T cells in 
Ref.~\refcite{harris}: while the speed $c$ is a function of CXCL10, other walk statistics are unaffected. 
Without explicit data stating otherwise CXCL10 is assumed here to be (approximately) uniformly 
distributed at the spatial scale of observed tissue, and hence $c$ is taken as spatially constant. Investigations into the impact of anti-CXCL10 treatments can be recreated through changing the 
size of $c$.

\subsection{Turn angle distribution}

To describe the motion of T cells we assume, following Ref.~\refcite{harris}, that the new 
direction is chosen independently of the target's position. Thus, we take
\begin{equation}
k(\mathbf{x},t,\mathbf{\theta};\mathbf{\eta})=\ell(\mathbf{x},t,|\eta-\theta|)\label{eq: turn angle distribution}
\end{equation}
where the new direction $\eta$ is symmetrically distributed with respect to the previous 
direction $\theta$, according to the symmetric distribution $\ell$. \cite{alt1980biased}  
$|\eta-\theta|$ denotes the distance between two directions on the unit sphere $S$.
More generally, immune cells can orient in response to environmental factors, such as 
attractant gradients or the structure of the extracellular matrix. In the
absence of data suggesting that such guidance cues play any (significant) role in
the behaviour observed in Ref.~\refcite{harris}, we presently exclude this possibility.

\subsection{Running probability and resting times}

As described in Ref.~\refcite{harris}, the motion of CD$8^+$ T cells is characterized by 
long runs, distributed according to a L\'{e}vy distribution, combined with resting 
times $r$. Within our microscopic description, we therefore assume the following 
power-law distribution for the running probability
\begin{equation}
    \psi(\mathbf{x},\tau)=\left(\frac{\tau_0(\mathbf{x})}{\tau_0(\mathbf{x})+\tau}\right)^\alpha\ , \ \textnormal{for}\ 1<\alpha<2 \label{eq: running probability}\ ,
\end{equation}
while resting times are distributed according to
\begin{equation}
    \psi_r(r)=\left(\frac{r_0}{r_0+r} \right)^\kappa \ \textnormal{for}\ 0<\kappa<1\ .\label{eq: resting probability}
\end{equation}
$\psi$ describes the probability that a moving cell stops after time $\tau$. The 
resting time distribution, $\psi_r$, gives the probability that a cell does not 
move for a time $r$. 

The running and waiting probabilities, $\psi$ and $\psi_r$, are related to the stopping and waiting frequency $\beta$ and $\beta_r$, via
\begin{align}
\psi(\mathbf{x},\tau) & =\exp\left(-\int_0^{\tau}\beta(\mathbf{x}+cs\theta,s) ds\right) \textnormal{and}\\ \psi_r(r) & =\exp\left(-\int_0^{r}\beta_r(s)ds \right)\ .
\end{align}
Moreover, explicit expressions for both rates, $\beta(\mathbf{x},\tau)$ and $\beta_r(r)$, can be computed from the relations:
\begin{align}
    \beta(\mathbf{x},\tau)& =\frac{\varphi(\mathbf{x},\tau)}{\psi(\mathbf{x},\tau)}=\frac{-\partial_\tau\psi}{\psi}=\frac{\alpha}{\tau_0+\tau}\ ,\label{eq: beta}\\ \beta_r(r)& =\frac{\phi(r)}{\psi_r(r)}=\frac{-\partial_r\psi_r}{\psi_r}=\frac{\kappa}{r_0+r}\label{eq: beta_r}\ .
\end{align}

\section{Modelling equations}\label{sec: modelling equations}

Considering the assumptions in Section \ref{sec: micro} and following the approach of Ref.~\refcite{alt1980biased}, 
densities of moving $\sigma(\mathbf{x},t,\theta,\tau)$ and resting $\sigma_0(\mathbf{x},t,\theta,\tau)$ populations 
are described by the following system of equations:
\begin{align}
    (\partial_\tau+\partial_t+c\theta\cdot\nabla)\sigma(\cdot,\theta,\tau)&=-\beta(\mathbf{x},\tau)\sigma(\cdot,\theta,\tau) \ , \label{eq: kinetic}\\
(\partial_t-\partial_\tau)\sigma_0(\cdot, \theta,\tau)& =T \beta(\mathbf{x},\tau)\sigma(\cdot,\theta,\tau)\ , \label{sigma0equation} \\ \sigma(\cdot,\theta,0)&= \sigma_0(\cdot,\theta,0)\ , \label{sigma0sigma}
\end{align}
where the dot denotes dependence in space, $\mathbf{x}$, and time $t$. Here the turn angle operator $T$, given by
\begin{equation}
T\phi(\eta)=\int_S k(\cdot,\mathbf{\theta};\mathbf{\eta})\phi(\theta)d\theta \label{eq: turn angle operator}\ ,
\end{equation} describes the effect of changing from direction $\mathbf{\theta}$ to a new direction $\mathbf{\eta}$. The initial condition for the particles that start a new run at $\tau=0$ is given by
\begin{equation}
    \sigma(\cdot,\eta,0)=\int_0^t\phi(r)\int_0^{t-r}d\tau\int_S\beta(\mathbf{x},\tau)\sigma(\mathbf{x},t-r,\theta,\tau)k(\cdot,\theta;\eta)d\theta dr\ .\label{eq: new run}
\end{equation}
The left hand side of equation \eqref{eq: kinetic} describes the temporal variation and transport of the density $\sigma(\cdot,\theta,\tau)$, while the right hand side gives the density of individuals that are left behind due to reorientation.  These particles reappear in the resting mode described by \eqref{sigma0equation}, where stopping with frequency $\beta({\mathbf{x},\tau})$ eventually generates a new run ($\tau=0$) following a pause of some time $r$, with a probability given by the probability density function $\phi(r)$. This is described by equations \eqref{sigma0sigma} and \eqref{eq: new run}. 

Using the method of characteristics we find the solution of (\ref{eq: kinetic}),
\begin{equation}
    \sigma(\cdot,\theta,\tau)=\sigma(\mathbf{x}-c\theta\tau,t-\tau,\theta,0)\exp\left(-\int_0^\tau\beta(\mathbf{x}+cs\theta,s)ds \right).\label{eq: solution}
\end{equation}
We can rewrite expression (\ref{eq: new run}) as
\begin{equation}
    \sigma(\cdot,\eta,0)=\int_Sk(\cdot,\theta;\eta)\left[\int_0^td\tau \int_0^{t-\tau}\phi(r)\beta(\mathbf{x},\tau)\sigma(\mathbf{x},t-r,\theta,\tau)dr\right]d\theta\ , \label{eq: initial run}
\end{equation}
after changing the limits of integration. Then, integrating (\ref{eq: kinetic}) and \eqref{sigma0equation} with respect to $\tau$ and substituting \eqref{sigma0sigma} and \eqref{eq: initial run}, we obtain
\begin{align}
    \partial_t\bar{\sigma}+c\theta\cdot\nabla\bar{\sigma} & =T\int_0^t\beta({\mathbf{x},\tau})\left(\int_0^{t-\tau}\phi(r)\sigma(\mathbf{x},t-r,\theta,\tau)dr\right)d\tau\nonumber\\ & -\int_0^t\beta(\mathbf{x},\tau)\sigma(\mathbf{x},t,\theta,\tau)d\tau\label{eq: tau independent}\ ,\\
    \partial_t\bar{\sigma}_0 &= T \int_0^t\beta(\mathbf{x},\tau)\sigma(\mathbf{x},t,\theta,\tau)d\tau\nonumber \\ & -T\int_0^t\beta({\mathbf{x},\tau})\left(\int_0^{t-\tau}\phi(r)\sigma(\mathbf{x},t-r,\theta,\tau)dr\right)d\tau\ . \label{eq: tau independent2}
\end{align}
Here $\bar{\sigma}$ and $\bar{\sigma}_0$ are defined as
\begin{equation}
\bar{\sigma}(\cdot,\theta)=\int_0^t\sigma(\cdot,\theta,\tau)d\tau,\ \ \bar{\sigma}_0(\cdot,\theta)=\int_0^t\sigma_0(\cdot,\theta,\tau)d\tau\ .\label{eq: sigma bar}
\end{equation}
From (\ref{eq: tau independent}) and (\ref{eq: tau independent2}) we can define the arrival rate of particles at a point $(\mathbf{x},t)$, after waiting for time $r$, as
\[
j(\cdot,\theta)=\int_0^t\beta(\mathbf{x},\tau)\left(\int_0^{t-\tau}\phi(r)\sigma(\mathbf{x},t-r,\theta,\tau)dr\right)d\tau
\]
and the density of cells leaving the point $\mathbf{x}$ for all times $\tau$ from $0$ to $t$, also called the escape rate, as
\begin{equation}
i(\cdot,\theta)=\int_0^t\beta(\mathbf{x},\tau)\sigma(\mathbf{x},t,\theta,\tau)d\tau\ . \label{eq: escape rate i}
\end{equation}
Using (\ref{eq: solution}) and the relations in (\ref{eq: beta}), we can write 
\[
i(\cdot,\theta)=\int_0^t\mathcal{B}(\mathbf{x},t-s)\bar{\sigma}(\mathbf{x}-c\theta(t-s),s,\theta)ds\ ,
\]
as derived in Ref.~\refcite{pks}, where $\mathcal{B}$ is given, in the Laplace space
\begin{equation}
\hat{\mathcal{B}}(\mathbf{x},\lambda+ c\theta\cdot\nabla)=\frac{\hat{\varphi}(\mathbf{x},\lambda+c\theta\cdot\nabla)}{\hat{\psi}(\mathbf{x},\lambda+c\theta\cdot\nabla)}+\textnormal{l.o.t.}.\label{eq:kernel}
\end{equation}

To rewrite $j(\cdot,\theta)$ in terms of $\bar{\sigma}$ we use (\ref{eq: solution}) again and 
let $s=t-\tau$. Hence,
\begin{align*}
    j(\cdot,\theta) & =\int_0^t\beta(\mathbf{x},\tau)\left(\int_0^{t-\tau}\phi(r)\sigma(\mathbf{x}-c\theta\tau,t-\tau-r,\theta,0)\psi(\mathbf{x},\tau)dr\right)d\tau\\ & =\int_0^t\beta(\mathbf{x},t-s) \psi(\mathbf{x},t-s)e^{-(t-s)c\theta\cdot\nabla}\left(\int_0^s\phi(s-r)\sigma(\mathbf{x},r,\theta,0)dr\right)ds\\ & =\int_0^t\varphi(\mathbf{x},t-s)e^{-(t-s)c\theta\cdot\nabla}\left(\phi(s)\ast\sigma(\mathbf{x},s,\theta,0) \right)ds\ .
\end{align*}
Note that $j(\cdot,\theta)$ is still written in terms of $\sigma$ instead of $\bar{\sigma}$. So, taking the Laplace transform of $j(\cdot,\theta)$ and using the relation,
\[
\hat{\bar{\sigma}}(\mathbf{x},\lambda,\theta)=\hat{\sigma}(\mathbf{x},\lambda,\theta,0)\hat{\psi}(\mathbf{x},\lambda+c\theta\cdot\nabla)
\]
which was obtained from (\ref{eq: sigma bar}) and (\ref{eq: solution}) (see Ref.~\refcite{pks} for details), we can rewrite as,
\begin{align*}
    \hat{j}(\mathbf{x},\lambda,\theta)=\hat{\varphi}(\mathbf{x},\lambda+c\theta\cdot\nabla)\hat{\phi}(\lambda)\hat{\sigma}(\mathbf{x},\lambda,\theta,0)=\frac{\hat{\varphi}(\mathbf{x},\lambda+c\theta\cdot\nabla)}{\hat{\psi}(\mathbf{x},\lambda+c\theta\cdot\nabla)}\hat{\phi}(\lambda)\hat{\bar{\sigma}}(\mathbf{x},\lambda,\theta)\ .
\end{align*}
Equations (\ref{eq: tau independent}) and (\ref{eq: tau independent2}) now can be written as
\begin{align}
    \partial_t\bar{\sigma}+c\theta\cdot\nabla\bar{\sigma} & =T\int_0^t\mathcal{B}(\mathbf{x},t-s)\left(\int_0^s\phi(s-r)\bar{\sigma}(\mathbf{x}-c\theta(t-s),r,\theta)dr\right)ds\nonumber\\ & -\int_0^t\mathcal{B}(\mathbf{x},t-s)\bar{\sigma}(\mathbf{x}-c\theta(t-s),s,\theta)ds\ ,\label{eq: important}\\ \partial_t\bar{\sigma}_0 & = T\int_0^t\mathcal{B}(\mathbf{x},t-s)\bar{\sigma}(\mathbf{x}-c\theta(t-s),s,\theta)ds\nonumber\\ & -T\int_0^t\mathcal{B}(\mathbf{x},t-s)\left(\int_0^s\phi(s-r)\bar{\sigma}(\mathbf{x}-c\theta(t-s),r,\theta)dr \right)ds\ .\label{eq: important 2}
\end{align}

\subsection{Scaling}\label{subsec: scaling}

Consider macroscopic space and time scales $\textsf{X}$ and $\textsf{T}$ respectively.
We assume that the mean run time $\bar{\tau}$ and the mean waiting time $\bar{r}$ are small compared to the macroscopic time, i.e. $\bar{\tau}/\textsf{T}$ and $\bar{r}/\textsf{T}$ are equal to $\varepsilon^{\textnormal{power}}\ll 1$. We scale as follows,
\begin{equation}
t_n= \varepsilon t\ ,\  \mathbf{x}_n=\frac{\varepsilon \mathbf{x}}{s}\ ,\  c_n=\varepsilon^{-\gamma} c_0\ , \ r_n=\varepsilon^\varrho r\ ,\ \textnormal{and}\  \tau_n=\tau\varepsilon^{\mu}\ ,\label{eq: scaling}
\end{equation}
for $\mu>0$, $\gamma>0$ and $\varrho>0$. \textcolor{black}{The scaling here is of parabolic type. It corresponds to a limit of the physical system with small average waiting and run times, small spatial run lengths, and large velocities compared to the macroscopic scales of an experiment. The values of the parameters $\gamma, \varrho, \mu$ are specified in Section \ref{section:finalequation}.}

Introducing this scaling we have,
\[
\psi_\varepsilon(\mathbf{x},\tau)=\left(\frac{\varepsilon^\mu\tau_0}{\varepsilon^\mu\tau_0+\tau} \right)^\alpha,\ \varphi_\varepsilon(\mathbf{x},\tau)=\frac{\alpha\left(\varepsilon^\mu\tau_0 \right)^\alpha}{\left(\varepsilon^\mu\tau_0+\tau\right)^{\alpha+1}}
\]
and 
\[\phi_{\varepsilon}(r)=\frac{\kappa\left(\varepsilon^\varrho r_0 \right)^\kappa}{(\varepsilon^\varrho r_0+r)^{\kappa+1}}\ .
\]
Moreover, (\ref{eq: important}) is given by
\begin{align}
    \varepsilon\partial_t\bar{\sigma}+\varepsilon^{1-\gamma}c_0\theta\cdot\nabla\bar{\sigma}& =T\int_0^t\mathcal{B}(\mathbf{x},t-s)\left(\int_0^s\phi_{\varepsilon}(s-r)\bar{\sigma}({\mathbf{x}-c\theta(t-s),r,\theta})dr\right)ds\nonumber\\ &-\int_0^t\mathcal{B}(\mathbf{x},t-s)\bar{\sigma}(\mathbf{x}-c\theta(t-s),s,\theta)ds\label{eq: important scaled}\ .
\end{align}
Computing the Laplace transform of the above expression we obtain 
\begin{align}
    \left(\varepsilon\lambda+\varepsilon^{1-\gamma}c_0\theta\cdot\nabla \right)\hat{\bar{\sigma}}&(\mathbf{x},\lambda,\theta) -\varepsilon\bar{\sigma}^0(\mathbf{x},\theta)\nonumber\\ & \simeq -\left(\mathds{1}-\hat{\phi}_{\varepsilon}\left(\varepsilon\lambda \right)T \right)\hat{\mathcal{B}}_\varepsilon\left(\mathbf{x},\varepsilon^{1-\gamma}c_0\theta\cdot\nabla \right)\hat{\bar{\sigma}}(\mathbf{x},\lambda,\theta)\ ,\label{eq: laplace space}
\end{align}
where we have assumed $\hat{\mathcal{B}}_\varepsilon(\mathbf{x},\varepsilon\lambda+\varepsilon^{1-\gamma}c_0\theta\cdot\nabla)\simeq\hat{\mathcal{B}}_\varepsilon(\mathbf{x},\varepsilon^{1-\gamma}c_0\theta\cdot\nabla)$ for $1>1-\gamma$.

The Laplace transform of the resting time density function $\phi_{\varepsilon}$, is given by
\[
\hat{\phi}_\varepsilon(\varepsilon\lambda)=\kappa\left(a\lambda \right)^\kappa\Gamma(-\kappa,a\lambda)e^{a\lambda}
\]
where $a=\varepsilon^{\varrho+1}r_0$. Using the following asymptotic expansion for the incomplete Gamma function  \cite{NIST:DLMF}
\begin{align}
\Gamma(b,z) & = \Gamma(b)\left( 1-z^{b}e^{-z}\sum_{k=0}^{\infty}\frac{z^k}{\Gamma(b+k+1)}\right)\ ,\label{eq: 3.23}
\end{align}
where $b$ is positive non-integer, and recalling that $b\Gamma(b)=\Gamma(b+1)$, we get
\begin{equation}
\hat{\phi}_\varepsilon(\varepsilon\lambda)=1-\varepsilon^{(1+\varrho)\kappa}r_0^\kappa\lambda^\kappa+\mathcal{O}(a\lambda)\label{eq: expasnion of phi}
\end{equation}
since $0<\kappa<1$. Note that in the above we have considered $e^{a\lambda}=1+\mathcal{O}(a\lambda)$ and this approximation is valid for $(1+\varrho)\kappa>0$.

Hence, substituting (\ref{eq: expasnion of phi}) into (\ref{eq: laplace space}) we obtain the following,
\begin{align}
    (\varepsilon\lambda+\varepsilon^{1-\gamma}&c_0\theta\cdot\nabla)\hat{\bar{\sigma}}(\mathbf{x},\lambda,\theta) -\varepsilon\bar{\sigma}^0(\mathbf{x},\theta)\nonumber\\ & \simeq -\left(\mathds{1}-(1-r_0^\kappa\varepsilon^{(1+\varrho)\kappa} \lambda^{\kappa})T\right)\hat{\mathcal{B}}_\varepsilon\left(\mathbf{x},\varepsilon^{1-\gamma}c_0\theta\cdot\nabla \right)\hat{\bar{\sigma}}(\mathbf{x},\lambda,\theta)\ .\label{eq: full Laplace transform}
\end{align}
Transforming back to the $(\mathbf{x},t)$-space we get
\begin{align}
    \varepsilon\partial_t\bar{\sigma}(\cdot,\theta)&+\varepsilon^{1-\gamma}c_0\theta \cdot\nabla\bar{\sigma}(\cdot,\theta) \simeq-(\mathds{1}-T)\mathcal{B}_\varepsilon(\mathbf{x},\varepsilon^{1-\gamma}c_0\theta\cdot\nabla)\bar{\sigma}(\cdot,\theta)\nonumber\\ & -r_0^\kappa\varepsilon^{(1+\varrho)\kappa}T{}_{t}\mathds{D}^{\kappa}\mathcal{B}_\varepsilon(\mathbf{x},\varepsilon^{1-\gamma}c_0\theta\cdot\nabla)\bar{\sigma}(\cdot,\theta)\ .\label{eq: important equation}
\end{align}
Here we have used the fact that the Laplace transform of the Riemann-Liouville fractional derivative ${}_{t}\mathds{D}^\kappa$ is given by Ref.~\refcite{klages2008anomalous}
\[
\mathcal{L}\left\{ {}_{t}\mathds{D}^\kappa f(t)\right\}=\lambda^\kappa \hat{f}(\lambda)- \sum_{m=0}^{n-1}\lambda^m\lim_{t\rightarrow 0}{}_{t}\mathds{D}^{\kappa-m-1}f(0^+)\ \textnormal{for}, \ n-1<\kappa<n\ ,
\]
where we assumed $f(0^+)=0$, since there is no scattering at time zero.

Scaling \eqref{eq: important 2} and changing the order of integration, the particles at rest satisfy the following equation
\begin{align}
    \varepsilon\partial_t\bar{\sigma}_0(\cdot,\theta)& =T\int_0^t\mathcal{B}(\mathbf{x},t-s)\bar{\sigma}(\mathbf{x}-c\theta(t-s),s,\theta)ds\nonumber \\ &-T\int_0^t\phi_\varepsilon(t-s)\left(\int_0^s\mathcal{B}(\mathbf{x},t-s')\bar{\sigma}(\mathbf{x}-c\theta(t-s'),s',\theta)ds'\right)ds\ .\label{eq: density particles at rest}
\end{align}
The Laplace transform of this expression is
\begin{align}
\varepsilon\lambda\hat{\bar{\sigma}}_0(\mathbf{x},\lambda,\theta)-\varepsilon\bar{\sigma}^0_0(\mathbf{x},\theta) =r_0^\kappa\varepsilon^{(1+\varrho)\kappa}\lambda^\kappa T\hat{\mathcal{B}}_\varepsilon(\mathbf{x},\varepsilon^{1-\gamma}c_0\theta\cdot\nabla)\hat{\bar{\sigma}}(\mathbf{x},\lambda,\theta)\ ,\label{eq: laplace transform resting particles}
\end{align}
and if we assume that $1>1-\gamma$ as before we get
\begin{align}
\varepsilon\partial_t\bar{\sigma}_0(\cdot,\theta) =r_0^\kappa\varepsilon^{(1+\varrho)\kappa}T{}_{t}\mathds{D}^\kappa\mathcal{B}_\varepsilon(\mathbf{x},\varepsilon^{1-\gamma}c_0\theta\cdot\nabla) & \bar{\sigma}(\cdot,\theta)\ .\label{eq: resting particles}
\end{align}

\subsection{Conservation of particles}\label{sec: resting particles}

From the system (\ref{eq: kinetic})-(\ref{sigma0sigma}) we can obtain a particle 
conservation equation, considering $\sigma_{\textnormal{tot}}(\mathbf{x},t,\theta)=\bar{\sigma}(\mathbf{x},t,\theta)+\bar{\sigma}_0(\mathbf{x},t,\theta)$, where $\bar{\sigma}$ and $\bar{\sigma}_0$ are given by (\ref{eq: important}) and (\ref{eq: important 2}) respectively. The conservation equation reads
\[
\varepsilon\partial_t\int_S\sigma_{\textnormal{tot}}d\theta+\varepsilon^{1-\gamma}c_0\int_S\theta\cdot\nabla\sigma_{\textnormal{tot}}d\theta=0\ ,
\]
where $S$ is the unit sphere.
Hence, substituting (\ref{eq: important equation}) and (\ref{eq: resting particles}) into the above expression we get
\begin{align}
 \varepsilon\partial_t\int_S\sigma_{\textnormal{tot}}d\theta &+\varepsilon^{1-\gamma}c_0\int_S\theta\cdot\nabla\sigma_{\textnormal{tot}}d\theta  =-\int_S(\mathds{1}-T)\mathcal{B}_\varepsilon(\mathbf{x},\varepsilon^{1-\gamma}c_0\theta\cdot\nabla)\bar{\sigma}d\theta\nonumber\\ &-r_0^\kappa\varepsilon^{(1+\varrho)\kappa}\int_S T\mathcal{B}_\varepsilon(\mathbf{x},\varepsilon^{1-\gamma}c_0\theta\cdot\nabla){}_{t}\mathds{D}^\kappa\bar{\sigma}(\mathbf{x},t,\theta)d\theta\nonumber\\ & +r_0^\kappa\varepsilon^{(1+\varrho)\kappa}\int_S T\mathcal{B}_\varepsilon(\mathbf{x},\varepsilon^{1-\gamma}c_0\theta\cdot\nabla){}_{t}\mathds{D}^\kappa\bar{\sigma}(\mathbf{x},t,\theta)d\theta=0\ . \label{eq: conservation long}
\end{align}
Note that here we have used the conservation of particles during the tumbling phase given in (\ref{eq: conservation T}).
If we consider $\sigma_{\textnormal{tot}}(\mathbf{x},t,\theta)=\frac{1}{|S|}\left(\bar{u}+\bar{u}_0+\varepsilon^\vartheta n\theta\cdot\bar{w}\right)$ then we finally have
\begin{equation}
    \varepsilon\partial_t(\bar{u}+\bar{u}_0)+\varepsilon^{\vartheta+1-\gamma} nc_0\nabla\cdot\bar{w}=0\ ,\label{eq: conservation}
\end{equation}
where
\[
\bar{u}_0(\mathbf{x},t)=\frac{1}{|S|}\int_S\bar{\sigma}_0(\cdot,\theta)d\theta\ ,
\]
and $\bar{u}$ and $\bar{w}$ are defined in Lemma \ref{lem: eigenfunctions}. The equation (\ref{eq: conservation}) is non-trivial only for $\vartheta=\gamma$.

We can define a new density, independent of the direction $\theta$, $u_{\textnormal{tot}}(\mathbf{x},t)=\bar{u}+\bar{u}_0$, that takes into account the moving and resting particles. Then, the conservation equation finally reads
\begin{equation}
    \partial_tu_{\textnormal{tot}}+nc_0\nabla\cdot\bar{w}=0\ .\label{eq: conservation eqaution}
\end{equation}

\section{Fractional space-time equation}\label{section:finalequation}

Next we obtain an expression for the mean direction $\bar{w}$, depending only on the density of moving particles $\bar{u}$.

Multiplying (\ref{eq: important equation}) by $\theta$ and integrating over all directions we obtain
\begin{align}
    n\varepsilon^{1+\gamma}\partial_t\bar{w}&+\varepsilon^{1-\gamma}c_0 \cdot\nabla\bar{u} \simeq-\frac{1}{|S|}\int_S\theta(\mathds{1}-T)\mathcal{B}_\varepsilon\left(\bar{u}+n\varepsilon^{\gamma}\theta\cdot\bar{w}\right)d\theta\nonumber\\ & -\frac{r_0^\kappa\varepsilon^{(1+\varrho)\kappa}}{|S|}\int_S\theta T\mathcal{B}_\varepsilon \ {}_{t}\mathds{D}^{\kappa}\left(\bar{u}+n\varepsilon^{\gamma}\theta\cdot\bar{w} \right)d\theta\ . \label{eq: expansion new densities}
\end{align}
From equation (\ref{eq:kernel}) and for $\hat{\varphi}(\mathbf{x},\varepsilon^{1-\gamma}c_0\theta\cdot\nabla)$ and $\hat{\psi}(\mathbf{x},\varepsilon^{1-\gamma}c_0\theta\cdot\nabla)$ given as in Ref.~\refcite{pks} we find
\begin{align}
\mathcal{B}_\varepsilon= \frac{\varepsilon^{-\mu}(\alpha-1)}{\tau_0}&-\frac{\varepsilon^{1-\gamma}c_0\theta\cdot\nabla}{2-\alpha}-\tau_0^{\alpha-2}\varepsilon^{\mu(\alpha-2)+(1-\gamma)(\alpha-1)}(1-\alpha)^2\nonumber\\ & \times
\Gamma(-\alpha+1)(c_0\theta\cdot\nabla)^{\alpha-1}  +\mathcal{O}\left(\tau_0^{\alpha-1}\varepsilon^{\mu(\alpha-1)}\lambda^\alpha \right)\ .
\end{align}
Substituting $\mathcal{B}_\varepsilon$ into (\ref{eq: expansion new densities}), we compare the leading powers of $\varepsilon$.
Considering that $(1+\varrho)\kappa>0$ as in Section \ref{subsec: scaling}, we observe that the terms involving the delay are of lower order with the exception of the term $\varepsilon^{-\mu+(1+\varrho)\kappa}{}_{t}\mathds{D}^\kappa\bar{u}$.  
The physically relevant scaling regime involves a fractional transport term in the expression for $\bar{w}$, hence we choose
\begin{equation}
\mu=\frac{1-\alpha(1-\gamma)}{\alpha-1}\ \textnormal{and}\ \gamma>1-\frac{1}{\alpha}\ \label{eq: scaling final} 
\end{equation}
to guarantee that $\mu>0$. Moreover, to ensure that the term involving a time delay is of lower order we also choose $(1+\varrho)\kappa>(\alpha-1)(\mu+1-\gamma)$. \textcolor{black}{Taking these relations into account, the right hand side of (\ref{eq: expansion new densities}) can be rewritten as 
\begin{align}
-\frac{1}{|S|}&\int_S\theta(\mathds{1}-T)\Bigl[\varepsilon^{-\mu}\frac{(\alpha-1)\bar{u}}{\tau_0}-\varepsilon^{-\mu+\gamma}\frac{(\alpha-1)}{\tau_0}n\theta\cdot\bar{w}-\tau_0^{\alpha-2}\varepsilon^{\mu(\alpha-2)+(1-\gamma)(\alpha-1)}\nonumber\\ &\times(1-\alpha)^2\Gamma(-\alpha+1)(c_0\theta\cdot\nabla)^{\alpha-1}\bar{u} \Bigr]d\theta+\mathcal{O}\left(\varepsilon^{\min\{-\mu+(1+\varrho)\kappa,\ \mu(\alpha-1)\}} \right).\label{eq: clearer}
\end{align}
From the coefficient 
of the leading term $\varepsilon^{-\mu}$ in (\ref{eq: clearer}) we then obtain,}
\begin{align}
    0=-\frac{1}{|S|}\int_S\theta(\mathds{1}-T)\frac{\alpha-1}{\tau_0}\bar{u}d\theta\label{eq: zero equation}\ .
\end{align}
The subleading term is of order $\varepsilon^{\mu(\alpha-2)+(1-\gamma)(\alpha-1)}$ and we get
\begin{align}
    0 & =-\frac{1}{|S|}\int_S\theta(\mathds{1}-T)\Bigl(\tau_0^{\alpha-2}(1-\alpha)^2\Gamma(-\alpha+1)c_0^{\alpha-1}(\theta\cdot\nabla)^{\alpha-1}\bar{u}\nonumber\\ & +\frac{n(\alpha-1)}{\tau_0}(\theta\cdot\bar{w}) \Bigr)d\theta\label{eq: first mean direction}\ .
\end{align}
Note that we have obtained the same fractional diffusion equation as in Ref.~\refcite{pks} and Ref.~\refcite{perthame2018fractional} 
for a constant chemoattractant concentration.

From (\ref{eq: first mean direction}) we can obtain the mean direction $\bar{w}$ after applying the operator $T$ to the right hand side. Therefore, we obtain
\begin{equation}
\bar{w}=\frac{\pi\tau_0^{\alpha-1}(\alpha-1)}{\sin(\pi\alpha)\Gamma(\alpha)}\frac{(n^2\nu_1-|S|)}{n|S|(\nu_1-1)}c_0^{\alpha-1}\nabla^{\alpha-1}\bar{u}\label{eq: mean direction final}\ .
\end{equation}
Substituting $\bar{w}$ into the conservation equation (\ref{eq: conservation eqaution}) we obtain
\begin{equation}
\partial_tu_{\textnormal{tot}}=\nabla\cdot \left(C_\alpha\nabla^{\alpha-1}\bar{u}\right)\label{eq: preliminary}\ ,
\end{equation}
where
\[
C_\alpha=-\frac{n\pi\tau_0^{\alpha-1}(\alpha-1)}{\sin(\pi\alpha)\Gamma(\alpha)}\frac{(n^2\nu_1-|S|)}{n|S|(\nu_1-1)}c_0^{\alpha}>0 \ \textnormal{for}\ 1<\alpha<2\ .
\]
Next we write the right hand side of (\ref{eq: preliminary}) in terms 
of $u_{\textnormal{tot}}$, and for this we return to the resting particles 
equation (\ref{eq: resting particles}).

Expanding the right hand side of (\ref{eq: resting particles}) and choosing only the 
leading terms we obtain, in the Laplace space,
\[
\lambda\hat{\bar{\sigma}}_0(\mathbf{x},\lambda,\theta)-\bar{\sigma}_0^0(\mathbf{x},0,\theta)=r_0^\kappa\lambda^\kappa \frac{(\alpha-1)}{\tau_0}\hat{\bar{u}}+\mathcal{O}\left(\varepsilon^{(1+\varrho)\kappa+\mu(\alpha-2)+(\alpha-1)(1-\gamma)} \right)\ .
\]
Here we have chosen $(1+\varrho)\kappa=\alpha(\mu+1-\gamma)$ which agrees with our previous assumption $(1+\varrho)\kappa>(\alpha-1)(\mu+1-\gamma)$. Integrating the above expression with respect to $\theta$ we can write it in terms of the Laplace transform of $\bar{u}_0$. Substituting $\bar{u}=u_{\textnormal{tot}}-\bar{u}_0$ into the right hand side and grouping terms we obtain
\begin{equation}
\hat{\bar{u}}_0(\mathbf{x},\lambda)-\frac{1}{\lambda}\bar{u}_0^0(\mathbf{x},0)=\frac{\hat{u}_{\textnormal{tot}}}{1+\frac{\tau_0}{r_0^\kappa(\alpha-1)}\lambda^{1-\kappa}}\ .
\end{equation}
Since $\lambda\rightarrow 0$ then, applying a Taylor expansion and assuming all particles are moving at $t=0$, i.e. $\bar{u}_0^0=0$, we have
\begin{equation}
\hat{\bar{u}}_0=\left(1-\frac{\tau_0\lambda^{1-\kappa}}{r_0^\kappa(\alpha-1)}+\mathcal{O}\left(\lambda^{2(1-\kappa)} \right) \right)\hat{u}_{\textnormal{tot}}\ .\label{eq: resting particles approximation}
\end{equation}
Substituting the inverse Laplace transform of (\ref{eq: resting particles approximation}) back into (\ref{eq: preliminary}) we get
\begin{align}
\partial_tu_{\textnormal{tot}} & =\nabla\cdot\left( C_\alpha\nabla^{\alpha-1}(u_{\textnormal{tot}}-\bar{u}_0)\right)\nonumber\\ &={}_{t}\mathds{D}^{1-\kappa} \nabla\cdot\left(C_{\alpha,\kappa}\nabla^{\alpha-1} u_{\textnormal{tot}} \right)\label{eq: FINAL}
\end{align}
where 
\[
C_{\alpha,\kappa}=\frac{\tau_0}{r_0^\kappa(\alpha-1)}C_\alpha\ .
\]
In fact, we can also write equation (\ref{eq: FINAL}) using the Laplace transform as
\begin{align}
     \lambda^{\kappa}\hat{u}_\textnormal{tot}-\lambda^{\kappa-1}u_\textnormal{tot}^0 & =\nabla\cdot\left(C_{\alpha,\kappa}\nabla^{\alpha-1}\hat{u}_\textnormal{tot}\right)\ ,
\end{align}
and using the fact that
\[
\mathcal{L}\left\{{}_{t}^C\mathds{D}^\kappa f(t) \right\}=\lambda^\kappa\hat{f}(\lambda)-\sum_{m=0}^{n-1}\lambda^{\kappa-m-1}f^{(m)}(0)\ ,
\] we have
\begin{equation}
{}_{t}^C\mathds{D}^\kappa  u_\textnormal{tot}=\nabla\cdot\left(C_{\alpha,\kappa}\nabla^{\alpha-1}u_\textnormal{tot} \right)\label{eq: final final}\ .
\end{equation}

 \begin{rem}
As previously noted, equation \eqref{eq: final final} does not contain a chemotactic component: this lies in agreement with Ref.~\refcite{harris}, where CD8$^+$  T cells do not exhibit directional migration on the time and length scales relevant to their experiments. 
 \end{rem}

\begin{rem}\label{sec: scaling relations}
From the analysis in the previous section, relevant scaling parameters satisfy 
the following relations:
\begin{align}
\mu & =\frac{1-\alpha(1-\gamma)}{\alpha-1}\ , \ \varrho =\frac{\alpha \gamma}{\kappa(\alpha-1)}-1\ ,
\end{align}
for $0<\kappa<1$ and $1<\alpha<2$. From (\ref{eq: scaling final}) and knowing that $\varrho>0$ we conclude that
\[
\kappa-\frac{\kappa}{\alpha}<\gamma<1-\frac{1}{\alpha} \ .
\]
For $\alpha=1.15$ and $\kappa=0.7$ as in \cite{harris}, $0.092<\gamma<1.15$. Choosing $\gamma=0.5$ then 
$$\mu\approx 3.8 \  \textnormal{and}\ \varrho\approx 4.47.$$ In this regime, the scaling of the long runs ($\mu$) and the scaling of the waiting times ($\varrho$) are of similar order.
\end{rem}

\subsection{Fundamental solution}\label{sec: fundamental solution}

Assuming that the stopping rate $\psi$ is independent of the position of the 
particle, we can write \eqref{eq: final final} as
\begin{equation}
{}_{t}^C\mathds{D}^\kappa\  u_\textnormal{tot}=C_{\alpha,\kappa}\nabla\cdot\left(\nabla^{\alpha-1}u_\textnormal{tot} \right)= \widetilde{C}_{\alpha,\kappa} (-\Delta)^{\alpha/2} u_\textnormal{tot} \label{eq: constant c}.
\end{equation}
Here, according to \eqref{appendixnormal} in two dimensions, for $1<\alpha<2$,
\[
\widetilde{C}_{\alpha,\kappa}=-2\sqrt{\pi} C_{\alpha,\kappa}\cos\left(\frac{\pi\alpha}{2} \right)\frac{\Gamma\left(\frac{\alpha+1}{2} \right)}{\Gamma\left(\frac{\alpha+2}{2} \right)}.
\]
Following Ref.~\refcite{fundsol}, the fundamental solution of \eqref{eq: constant c} in $\mathbb{R}^n$, with initial condition $\delta_0$ and diffusion constant $C_{\alpha,\kappa}$,, can be found with the help of the Fourier-Laplace transform
\begin{equation}
\hat{\overline{G}}(\lambda, \xi) = \frac{\lambda^{\kappa-1}}{\lambda^{\kappa} + \widetilde{C}_{\alpha,\kappa} \vert \xi \vert^{\alpha}}.
\end{equation}
Note that the Laplace transform of the Mittag-Leffler function is
\begin{equation}
\mathcal{L}\; E_{\kappa} (ct^{\kappa}) =  \frac{\lambda^{\kappa-1}}{\lambda^{\kappa} -c}.
\end{equation}
Thus,
\begin{equation}
\hat{G}(t,\xi) = E_{\kappa} ( -\widetilde{C}_{\alpha,\kappa} \vert \xi \vert^{\alpha} t^{\kappa}).
\end{equation}
Using the formula for the inverse transformation of a radial function, we obtain
\begin{equation}
G(t,\mathbf{x}) = \frac{\vert \mathbf{x} \vert^{1-n/2}}{(2\pi)^{n/2}} \int_0^{\infty} E_{\kappa} (-\widetilde{C}_{\alpha,\kappa}\tau^{\alpha} t^{\kappa}) \tau^{n/2} J_{n/2-1}(\tau \vert \mathbf{x} \vert) d \tau,
\end{equation}
where $J_r(z)$ is a Bessel function. Passing through the Mellin and inverse Mellin transform we conclude
\begin{equation}
G(t,\mathbf{x}) = \frac{1}{\pi^{n/2}|\mathbf{x}|^{n}} H^{2,1}_{2,3}\left(\frac{|\mathbf{x}|^{\alpha}}{2^{\alpha} \widetilde{C}_{\alpha,\kappa} t^\kappa} \Big|^{(1,1); (1, \kappa)}_{(n/2,\alpha/2); (1, 1);(1,\alpha/2)}\right)\label{eq: asymptotic of fundamental}\ ,
\end{equation}
where $H^{2,1}_{2,3}(z)$ is a Fox $H$-function. Useful identities and asymptotics may be found in Ref.~\refcite{Braaksma} and Ref.~\refcite{Hbook}. In particular by Theorem 3 in Ref.~\refcite{Braaksma} for $1< \alpha <2,\; 0<\kappa<1,\;\mathrm{and}\; \alpha<2\kappa$, 
\begin{equation}\label{eq: asympt}
G(t,\mathbf{x}) \simeq \frac{1}{|\mathbf{x}|^{n}} \ \left(\frac{|\mathbf{x}|^{\alpha}}{2^{\alpha} \widetilde{C}_{\alpha,\kappa} t^\kappa}\right)^q
\end{equation}
when $\frac{|\mathbf{x}|^{\alpha}}{\widetilde{C}_{\alpha,\kappa} t^\kappa} \ll 1$, where $q= 1$.
In the limit $\frac{|\mathbf{x}|^{\alpha}}{\widetilde{C}_{\alpha,\kappa} t^\kappa} \gg 1$, we have $q = -1$. Note that these estimates hold in the regime of the experiments in Ref.~\refcite{harris} discussed above, as well as for examples of superdiffusion without waiting times, \cite{pks} relevant for certain studies of \textit{E.~coli} and \textit{Dictyostelium discoideum}.\\

Other regimes generate the exponentially small tails known for Brownian motion, including the presence of waiting times. For example, for Brownian motion with waiting times, 
corresponding to $\alpha = 2$ and $\kappa<1$, we obtain the fundamental solution
\begin{equation}
G(t,\mathbf{x}) = \frac{1}{2 \pi^{n/2}|\mathbf{x}|^{n}} H^{2,0}_{1,2}\left(\frac{|\mathbf{x}|}{2 \sqrt{\widetilde{C}_{1,\kappa}} t^{\kappa/2}} \Big|^{(1,\kappa/2)}_{(1,1/2); (n/2, 1/2)}\right)\ .
\end{equation}
It has exponentially small tails as $|\mathbf{x}| t^{-\kappa/2} \to \infty$:
\begin{equation}
G(t,\mathbf{x}) \simeq \frac{1}{|\mathbf{x}|^n}\left(\frac{|\mathbf{x}|}{2 \sqrt{\widetilde{C}_{2, \kappa}} t^{\kappa/2}} \right)^{-\frac{n}{2-\kappa}} \exp \left( 2(\frac{\kappa}{2}-1) \kappa^{\frac{\kappa}{2-\kappa}}\left(\frac{|\mathbf{x}|}{2 \sqrt{\widetilde{C}_{2, \kappa}} t^{\kappa/2}}\right)^{\frac{2}{2-\kappa}}\right).
\end{equation}
 In particular, the  range in which the asymptotics \eqref{eq: asympt} holds shrinks to $0$ when $\alpha$ and $\kappa$ approach the boundary of the admissible region $1< \alpha <2,\; 0<\kappa<1,\;\mathrm{and}\; \alpha<2\kappa$. 

\subsection{Hitting times}

The fundamental solution of the continuum model derived in the previous subsection allows us 
to extract analytical approximations for biologically relevant quantities. As an example, we derive an expression for the time at which a particle hits some distant target $T$ with 
radius $a$, in the experimentally relevant regime $1< \alpha <2,\; 0<\kappa<1,\;\mathrm{and}\; \alpha<2\kappa$. We seek the first time at which the density of the solution in $T$ reaches a certain threshold $\delta$. That is, we seek $t_0$ such that
\begin{equation}
\delta = \int_T \int_{\mathbf{R}^n}  G(\mathbf{x}-\mathbf{y},t_0)u_0(\mathbf{y}) d\mathbf{y} d\mathbf{x}.
\end{equation}
Assuming that the initial positions of the particles are given by $\mathbf{x}_i$, so that $u_0(\mathbf{x})  = \sum_i \delta_{\mathbf{x}_i}(\mathbf{x})$, we obtain
\begin{align}
\delta &= \sum_i \int_T G(\mathbf{x}-\mathbf{x}_i,t_0) dx \nonumber \\& = \sum_i \int_T \frac{1}{\pi^{n/2}|\mathbf{x}-\mathbf{x}_i|^{n}} H^{2,1}_{2,3}\left(\frac{|\mathbf{x}-\mathbf{x}_i|^{\alpha}}{2^{\alpha} \widetilde{C}_{\alpha,\kappa} t_0^\kappa} \Big|^{(1,1); (1, \kappa)}_{(n/2,\alpha/2); (1, 1);(1,\alpha/2)}\right)\ d\mathbf{x}
\end{align}
If all initial positions are at distance $\gg (\widetilde{C}_{\alpha,\kappa} \tau^\kappa)^{1/\alpha}$ from the target $T$, we may use the asymptotic expansion of the $H$-function from the previous subsection to obtain
\begin{equation}
\begin{split}
\delta &\simeq \frac{2^{\alpha} \widetilde{C}_{\alpha,\kappa} t_0^{\kappa}}{\pi^{n/2}} \sum_i \int_T \vert \mathbf{x}-\mathbf{x}_i \vert^{-\alpha - n} d\mathbf{x} \\
& \simeq \frac{2^{\alpha} \widetilde{C}_{\alpha,\kappa} t_0^{\kappa}}{\pi^{n/2}} \mathrm{vol}(T) \sum_i \vert \mathbf{x}_0 - \mathbf{x}_i\vert^{-\alpha-n},
\end{split}
\end{equation}
where $\mathbf{x}_0$ is a centre of the target $T$. Thus,
\begin{equation}\label{hittimes}
t_0 \simeq \left( \frac{\delta \pi^{n/2}}{2^{\alpha} \widetilde{C}_{\alpha,\kappa} \mathrm{vol}(T) \sum_i \vert \mathbf{x}_0 - \mathbf{x}_i \vert^{-\alpha-n}} \right)^{1/\kappa}.
\end{equation}
This formula holds in the regime where the asymptotic expansion \eqref{eq: asympt} is valid. 

\section{Numerical methods} \label{numerics}

In addition to the detailed qualitative information provided by the fundamental solution, 
the space-time fractional continuum equation allows efficient quantitative 
modelling of immune cell behaviour. We briefly describe the numerical approximation of the 
nonlocal operators. Challenges include the numerical evaluation of the singular 
integrals and the lack of boundary regularity, which leads to reduced convergence 
rates in naive approaches. Our numerical approximation of Equation \eqref{eq: final final} 
uses a finite element discretisation in space as discussed, for example, in Ref.~\refcite{VIPreprint} 
and a time stepping method based on convolution quadrature as in Ref.~\refcite{acosta2017finite}.

Let $\Omega \subset \mathbb{R}^n$ be a bounded domain with polygonal boundary and let $f\in C^0([0,T) \times \Omega)$. For $\alpha \in (1,2)$ and $\kappa \in (0,1)$, we consider the problem
\begin{align}\label{eq: num_set}
{}_{t}^C\mathds{D}^\kappa u + \nabla \cdot (C_{\alpha,\kappa} \nabla^{\alpha-1}  u) & = f & \mathrm{in}\; \Omega \times [0,T) \nonumber\\
u &= 0 & \mathrm{in}\; \Omega^c \times [0,T)  \\ 
u(\cdot, 0) &= u_0 & \mathrm{in}\; \Omega. \nonumber
\end{align} 
Let $\mathcal{T}_h$ be a shape regular and quasi-uniform triangulation of the region $\Omega$, with triangles of diameter at most $h$. Let $H_h$ be the subspace of piecewise linear functions of $H_0^{\alpha/2}(\Omega)$ associated with $\mathcal{T}_h$. Then, the semidiscrete weak formulation of the problem is as follows: Find $u_h \in C^0([0,T); H_h) \cap C^\kappa([0,T); L^2(\Omega))$ such that 
\begin{align}
({}_{t}^C\mathds{D}^\kappa u_h,v) + a(u_h,v)  &= (f,v),\\
u_h(0) &= u_{0},
\end{align}
for all $v \in H_h$. Here $a(\cdot,\cdot)$ represents the bilinear form 
$$a(u,v) = (C_{\alpha,\kappa} \nabla^{\alpha-1} u, \nabla v)$$
of the fractional Laplacian and, for simplicity, we assume that $u_0 \in H_h$.
The discrete fractional Laplace operator $\Lambda_h$ is defined as the unique operator satisfying 
\begin{equation}
(\Lambda_h u_h,v_h) = a(u_h,v_h),\; \mathrm{for \; all}\; u_h,v_h \in H_h,
\end{equation}
and the mass matrix $M_h$ is given by 
\begin{equation}
(M_h u_h,v_h) = (u_h,v_h),\; \mathrm{for \; all}\; u_h,v_h \in H_h.
\end{equation}
We conclude a strong reformulation of the semidiscrete problem: Find $u_h \in C^0([0,T); H_h) \cap C^\kappa([0,T); L^2(\Omega))$ such that 
\begin{align}
 M_h {}_{t}^C\mathds{D}^\kappa u_h + \Lambda_h u_h  &= f_h & \mathrm{in}\; \Omega \times [0,T) \\
u_h &= 0 & \mathrm{in}\; \Omega^c \times [0,T) \nonumber\\ 
u(\cdot, 0) &= u_0 & \mathrm{in}\; \Omega\ .\nonumber 
\end{align}
For the discretisation of this equation in time we follow the approach of Ref.~\refcite{acosta2017finite}. Dividing the time interval $[0,T)$ uniformly with time step $\tau = T/N$ of size $h^{\alpha/\kappa}$, we seek a numerical approximation of the convolution integral $K \ast g(t)$ associated with the Caputo time fractional derivative, by means of a finite sum as
\begin{equation}
K \ast g(t) = \int_0^t K(s)g(t-s) \mathrm{d}s \approx \sum_{j=0}^n w_j g(t-j\tau).
\end{equation} 
The weights $w_j$ are computed from the Taylor expansion of $\mathcal{K}(\delta(y)/\tau)$. Here, $\mathcal{K}$ is the Laplace transform of the kernel $K$ and $\delta(y)$ is the quotient of the generating polynomials of a multistep method. 
The $w_j$ are calculated from the recursion relation
\begin{align*}
w_0 &= \tau^{-\alpha/2},\\
w_j &= \left(1-\frac{\alpha+2}{2 j}\right)w_{j-1}.
\end{align*}
For full details see Ref.~\refcite{Lubich1988_1} and Ref.~\refcite{Lubich_2}. \\
The fully discrete time stepping scheme for \eqref{eq: num_set} is then given as follows: Find $\{u_h^1, u_h^2, \dots \} \subset H_h$ such that
\begin{equation}
(w_0 M + A) u_h^n = M \left((\sum_{j=0}^n w_j) u_h^0 - \sum_{j=1}^n w_j u_h^{n-j} + f_h^n \right),
\end{equation}
where $u_h^0$ is given. $M, A$ are the mass and stiffness matrices related to the piecewise linear basis functions $\varphi_i$ of $H_h$ defined by $M_{ij} = (\varphi_i,\varphi_j),\; A_{ij} = a(\varphi_i,\varphi_j)$, and $f_h^n = \sum_i (f(\cdot, n\tau), \phi_i) \phi_i$ is the Galerkin projection of $f$ onto $H_h$ at time $n \tau$.\\

\begin{figure}[!ht]
\centering
\subfloat[][]{
\includegraphics[width = 0.5\textwidth]{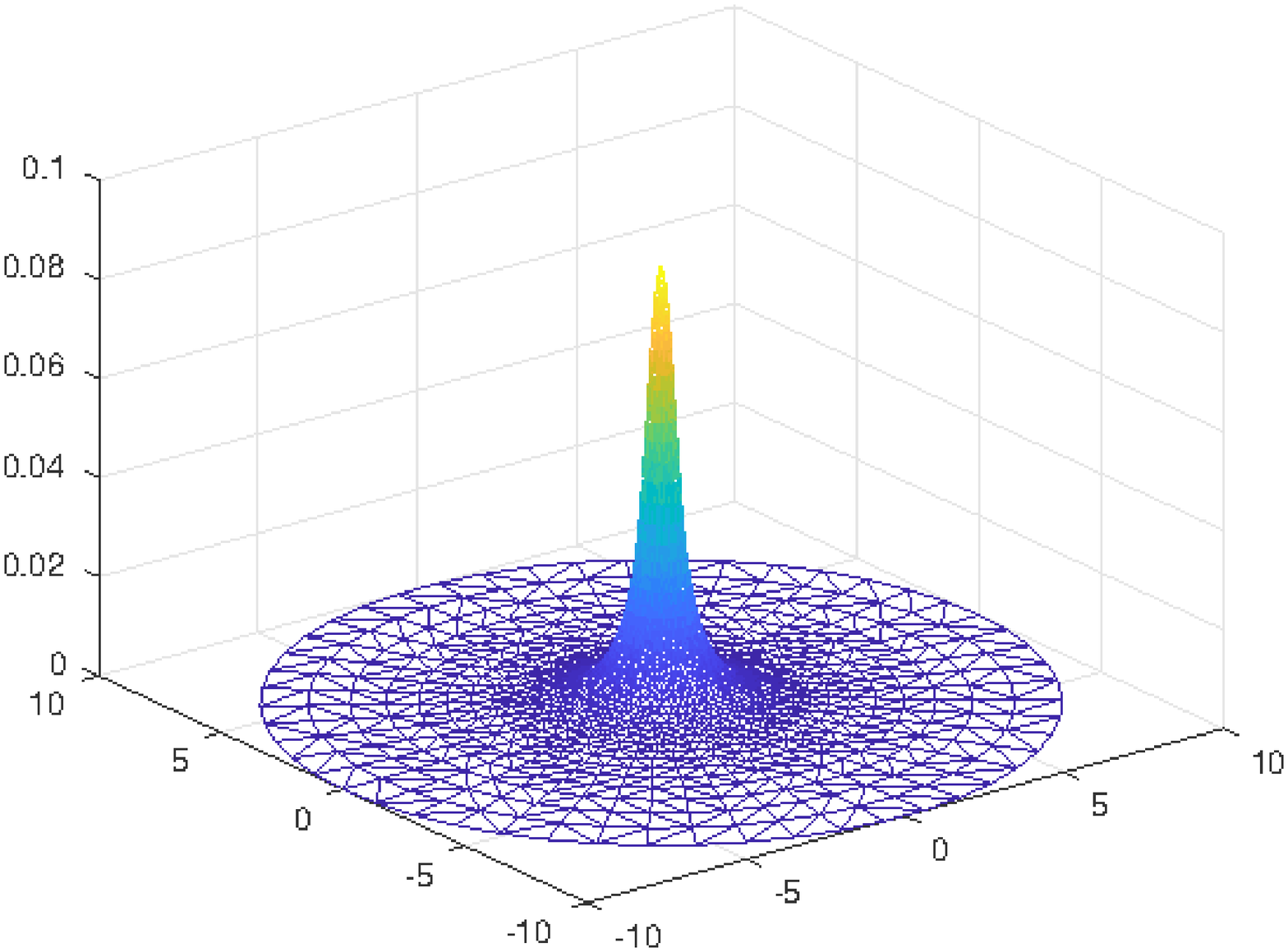}
\label{f:delay}
}
\subfloat[][]{
\includegraphics[width = 0.5\textwidth]{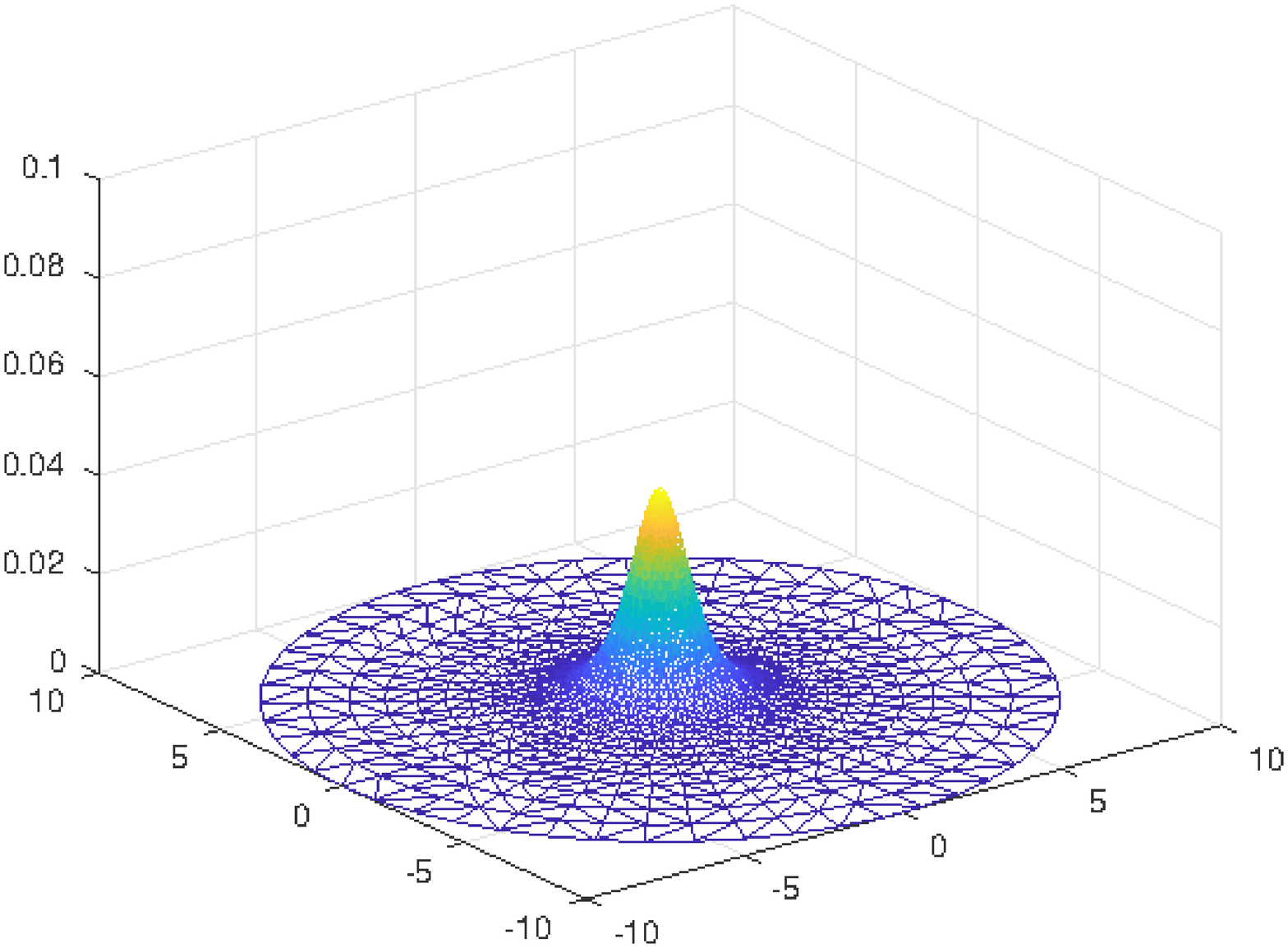}
\label{f:nodelay}
}
\caption{\textbf{(a)} Solution to \eqref{eq: num_set} at time $t = 1$ for $\alpha = 1.15$, $\kappa = 0.7$ (resting). \textbf{(b)} Solution to \eqref{eq: num_set} at time $t = 1$ for $\alpha = 1.15$, $\kappa = 1$ (no resting).}
\end{figure}

\begin{figure}[!ht]
\centering
\includegraphics[width = \textwidth]{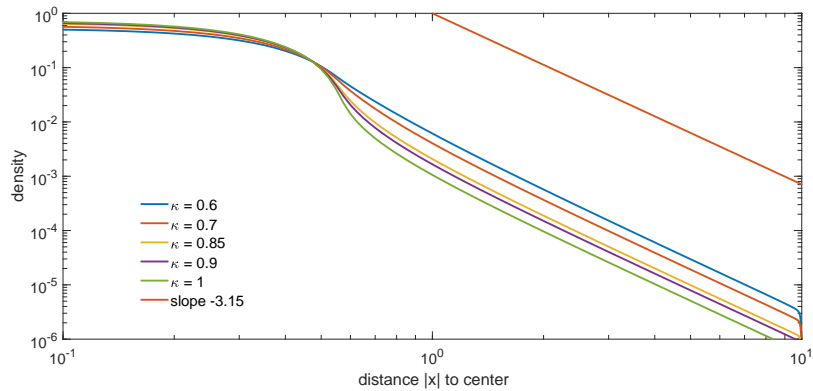}
\caption{Cross-section of solution depending on $\kappa$ for $\alpha = 1.15$.}
\label{f:A107Kvar}
\end{figure}

\begin{figure}[!ht]
\centering
\includegraphics[width = \textwidth]{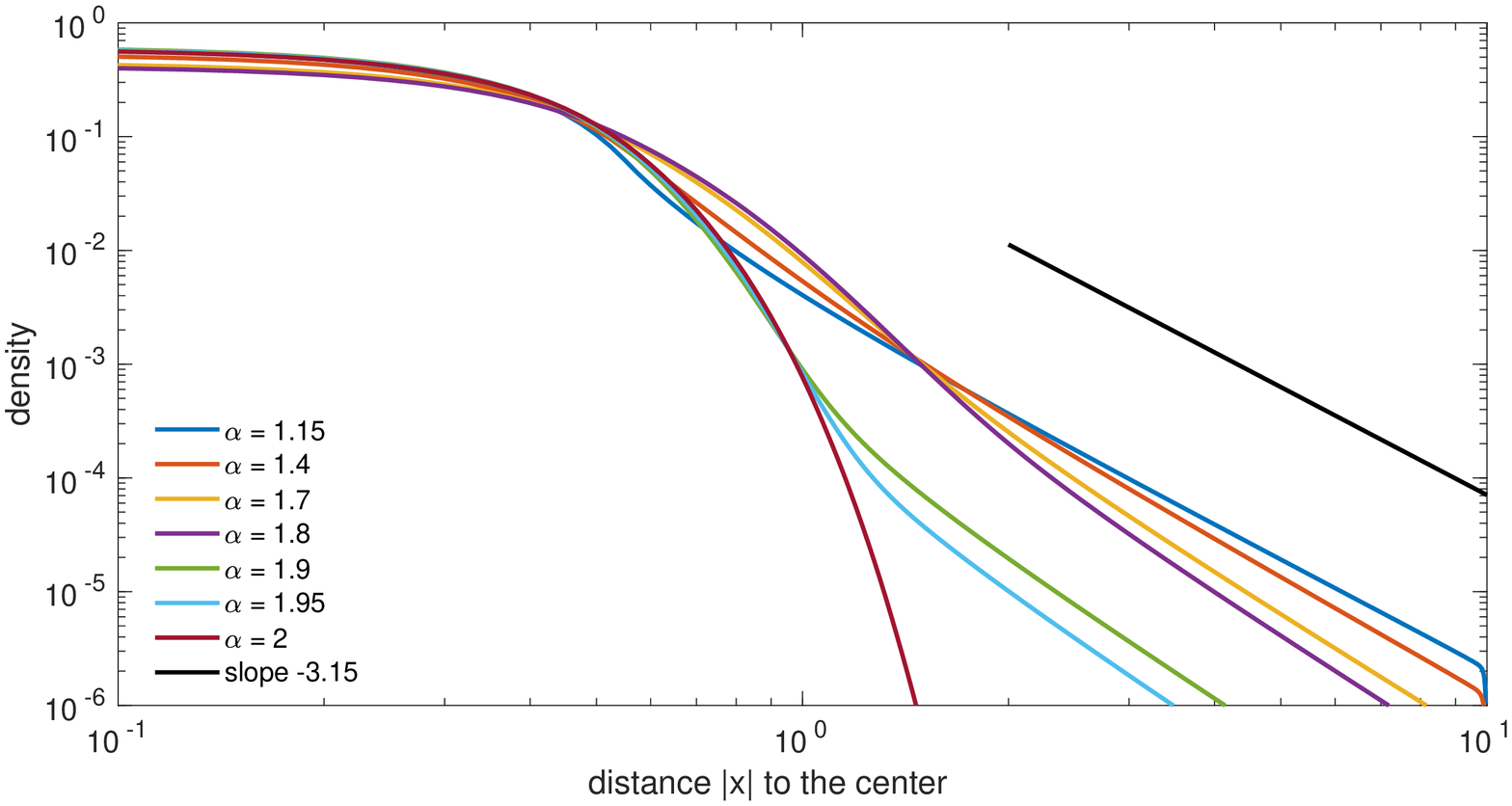}
\caption{Cross-section of solution depending on $\alpha$ for $\kappa = 0.7$.}
\label{f:K07Avar}
\end{figure}

To illustrate the effect of the fractional derivative in time in the biologically relevant regime, we consider problem \eqref{eq: num_set} in (a polygonal approximation of) $\Omega = B(0,10)$ with $f \equiv 0 $ and $u_0(\mathbf{x}) = \mathrm{max}(\exp(-5 \vert \mathbf{x} \vert^2) - 0.2, 0)$, using  $h \simeq 0.025$ and $\tau \simeq h^{\alpha/\kappa}$. \textcolor{black}{This setup corresponds to a Petri dish with an initial density of cells in the center. The domain is large enough so that the dominant effects correspond to diffusion rather than  boundary effects, as in the experiment \cite{harris}.} The solution at time $t = 1$ is shown for $\alpha = 1.15$ and $\kappa = 0.7$ in Figure \ref{f:delay} and for $\alpha = 1.15$ and $\kappa=1$ in Figure \ref{f:nodelay}. The figures clearly exhibit the memory effects induced by the fractional derivative in time.
\begin{figure}[!ht]
\centering
\includegraphics[width = \textwidth]{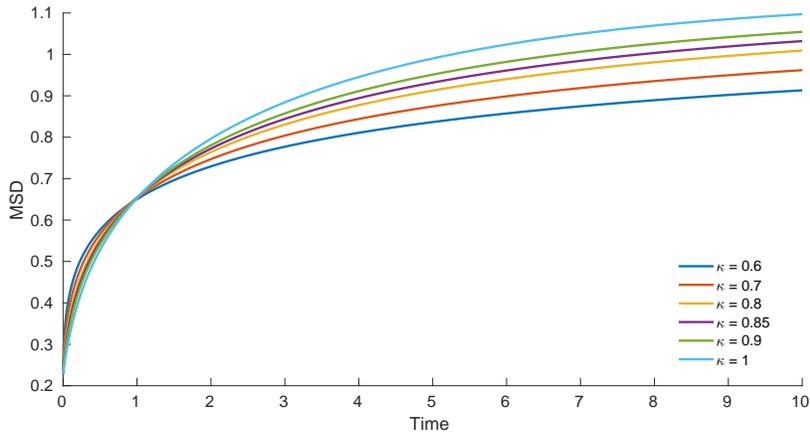}
\caption{Width of solution depending on $\kappa$ as a function of time for $\alpha = 1.15$.}\label{f:CrossPlot}
\end{figure}
\textcolor{black}{Figure~\ref{f:CrossPlot} shows the width of the cell density as a function of time for different $\kappa$.}\\
Cross-sections of solutions of Equation~\eqref{eq: final final} with initial condition given by $u(0,\mathbf{x}) = \mathrm{max}(\exp(-5 \vert \mathbf{x} \vert^2) - 0.2, 0)$ are given in Figures~\ref{f:A107Kvar} and \ref{f:K07Avar}, for time $t=0.02$.
\textcolor{black}{The time $t = 0.02$ is chosen in order to exhibit the long tails of the L\'{e}vy diffusion, with their known slope. The influence  of the boundary becomes more relevant for longer times.}
 Figure~\ref{f:A107Kvar} shows a cross section of the solution for values of $\kappa$ from $0.6$ to $1$, for the experimentally obtained $\alpha=1.15$ as in \cite{harris}. In particular, it depicts the expected tail of the density with decay $|\mathbf{x}|^{-n-\alpha} = |\mathbf{x}|^{-3.15}$, independent of $\kappa$, as well as the Markovian limit $\kappa \to 1$. Figure~\ref{f:K07Avar} varies the coefficient $\alpha$ from $1.15$ to $2$, for $\kappa=0.7$ as in Ref.~\refcite{harris}. As long as $\alpha<2$, the density again decays like $|\mathbf{x}|^{-n-\alpha}$ away from the initial bump, while it exhibits the faster Gaussian decay for $\alpha = 2$. \textcolor{black}{As $\alpha \to 2^-$ the onset of algebraic decay is only visible on larger and larger spatial scales. We highlight the fact that the exponent of the decay does not depend on $\kappa$. This is due to the fact that the decay exponent of the fundamental solution for  $|x| \to \infty$ depends only on $\alpha$, while it is independent of $\kappa$, see \eqref{eq: asympt}.}

\section{Conclusions \& Outlook}

In this paper we have derived effective macroscopic diffusion equations for organisms 
exhibiting long-range behaviour and pauses. Beginning with a microscopic model in which 
run times and waiting times followed a power-law, as observed for certain T cell populations controlling 
chronic infections, we obtained a system of kinetic equations for the moving and resting 
particles. The fractional diffusion equation \eqref{eq: final final} emerges in a 
realistic limit.

The paper initiates a study into the interplay between long-range behaviour in 
space and long delays between runs, contributing to recent interest in 
anomalous diffusion processes. On the one hand, L\'{e}vy walks in space with 
short / negligible delays have been suggested for the movements of organisms 
such as \textit{E.~coli} under low nutrient levels and their macroscopic evolution has been
shown to be described by fractional Patlak-Keller-Segel equations. 
\cite{bellouquid2016kinetic, pks, perthame2018fractional} They have also inspired search strategies for swarm robotic systems.\cite{RoboticsPreprint} On the other hand, Brownian 
motion with subdiffusive behaviour in time has been investigated in the 
context of death processes \cite{fedotov2015persistent} or nonlinear interactions. \cite{edotov2013nonlinear, straka2015transport} A discussion of resting 
times in velocity-jump models is found in Ref.~\refcite{TaylorKing}. 

The macroscopic diffusion equation \eqref{eq: final final} permits analytical 
insights into the evolution of the density. For example, it reveals that the microscopic 
description enters via three parameters: the exponents $\alpha$ and $\kappa$ of the run and 
waiting times and the diffusion constant $C_{\alpha,\kappa}$. Chemotactic terms are of
lower order: the long-range searching strategy is thus not disrupted by local 
gradient following. Of course, immune cells are well known for their responses
to chemoattractants \cite{griffith2014chemokines}: in the context of the T cells studied here, 
it is possible that their detection of a local attractant gradient would trigger a conversion from
long range searching behaviour to local gradient following.

The fundamental solution in $\mathbb{R}^n$, \eqref{eq: asymptotic of fundamental},
provides an explicit formula for the probability distribution for the movement of a 
single particle. It leads to approximations for hitting times, \eqref{hittimes}, allows us 
to study the sensitivity to parameter changes and provides a step towards the analysis 
of mean first passage times, see below.  On the other hand, Section \ref{numerics} offers efficient and accurate numerical methods to employ the fractional PDE \eqref{eq: final final} for parametric studies, despite its nonlocal nature, and more extensive modelling is 
addressed elsewhere.

\begin{figure}[!ht]
\centering
\includegraphics[width = 0.8\textwidth]{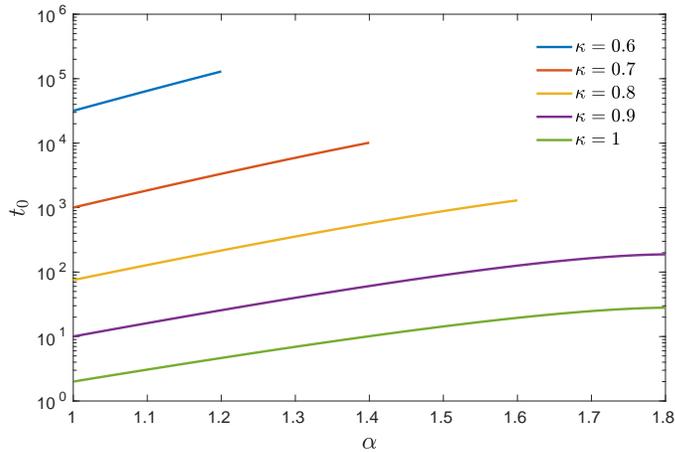}
\caption{Hitting time $t_0$ as function of $\alpha$ and $\kappa$ in the range of validity of \eqref{hittimes}.} \label{alphakappa}
\end{figure}

The experiments of Ref.~\refcite{harris} specifically studied the effect of the CXCL10 concentration 
on T cell velocity: CD8$^+$ T cells in mice treated with anti-CXCL10 were, on average, 23\% 
slower than the cells of a control population with normal responses. From the 
fundamental solution of the space-time fractional equation we observe that velocity 
changes only alter the time scale of diffusion, corresponding to $c^{\frac{\alpha}{\kappa}}$. 
Thus, for the experimentally determined values $\alpha = 1.15$ and $\kappa = 0.7$, a 23\% reduction in the velocity would yield an approximately 35\% reduction in the diffusion timescale, and hence less efficient searching. In the absence of data stating otherwise, 
here we have assumed the CD8$^+$ T cells migrate in an environment with 
homogeneous CXCL10 levels, and therefore constant velocities $c$. More generally, it
would be of high interest to explore the impact of spatially-dependent velocities, resulting 
from nonuniform chemical profiles. The microscopic modelling of such problems, however, 
appears to be challenging even for velocity-jump models with standard Brownian motion. 
The model could also be extended to include extra complexity. For example, in bacteria, 
the stopping probability is linked to molecular components, which could enter as internal variables. 
We refer to work in this direction by Perthame et~al.~\cite{perthame2018fractional} for the run-and-tumble of 
bacteria including a biochemical pathway, and a more detailed discussion of the impact of 
including internal variables is provided in. \cite{xue2009individual}  

In the context of organisms searching for targets, a basic quantity of interest is the mean first passage time. It is defined as the time taken for a moving organism to reach a target or, more formally \cite{gardiner1986handbook} as
\[
\mathcal{T}(\mathbf{x})=\int_0^\infty \int_{\Omega}p(\mathbf{x}',t\mid \mathbf{x},0)d\mathbf{x}'\ dt\ .
\]
Here, $p(\mathbf{x}',t\mid \mathbf{x},0)$ is the probability that the particle is at $\mathbf{x}'$ at time $t$ provided that it was at $\mathbf{x}$ at time $0$, i.e.~the Green's function of the fractional equation. For the diffusion equation \eqref{eq: final final} two regimes have been considered in one dimension: For subdiffusion, $0<\kappa<1$ and $\alpha=2$, it was shown in Ref.~\refcite{yuste2004comment} that $\mathcal{T}(\mathbf{x})\rightarrow\infty$ for a target in a bounded domain, while for superdiffusion, $\kappa=1$ and $1<\alpha<2$, $\mathcal{T}(\mathbf{x})$ is finite under the same conditions. \cite{gitterman2000mean} It is the nonlocality of the equation here that generates the challenge, raising as it does the possibility of ``leapfrogging'' a target. The analysis in higher dimensions remains open. \cite{yuste2004comment}

\textcolor{black}{Motivated by the differential movement of cells in gray and white brain matter,\cite{giese1996migration}
upcoming work on interface problems will consider velocities that take different values
in distinct regions of the domain. While our current article addresses uncorrelated
run and waiting times, correlations between these are also of interest. In the special
case of perfect correlations between run and waiting times the macroscopic limit
coincides with the one obtained from a velocity jump model for a correspondingly
reduced velocity. Weaker forms of correlation are an interesting topic for future
research.}

From a search-area coverage perspective, a long-tailed distribution of waiting times makes 
little sense: Figure \ref{alphakappa} shows that waiting only increases the hitting time, and hence 
decreases the searching efficiency. Of course, such apparent contradictions can only be explained
through considering the underlying problem: following a migration, T cells must spend a certain time 
controlling their local environment for any antigen presenting (i.e. infected) cells, often 
detected through direct cell-cell contact, and hence `waiting' is an intrinsic component 
of the search/detection process. While we have followed the data of Ref.~\refcite{harris} and assumed
independence between the selection of run and wait times, it is of course possible that a link
exists: for example, a T cell performs a thorough check of some environment (checks a large number of 
cells) before embarking on a long run. The extent to which such considerations impact on the 
subsequent PDE remain to be explored.

\appendix

\section{Turn angle and fractional operators}\label{sec: operator T}
This section specifies some basic spectral properties of the turn angle operator $T$ defined in (\ref{eq: turn angle operator}).
Because $\ell$ in (\ref{eq: turn angle distribution}) is a probability distribution, it is normalized to $\int_S \ell(\mathbf{x},t,|\theta-e_1|)d\theta=1$,  where $e_1 = (1,0,\dots, 0)$. We immediately observe
\begin{align}
    \int_S (\mathds{1}-T)\phi d\theta=0\label{eq: conservation T}
\end{align}
for all $\phi\in L^2(S)$. Biologically, (\ref{eq: conservation T}) corresponds to the conservation of the number of organisms in the tumbling phase.
We also require some more detailed information about the spectrum of $T$. 

\begin{lem}\label{lem: eigenfunctions}
Assume that $\ell$ is continuous. Then $T$ is a symmetric compact operator. In particular, there exists an orthonormal basis of $L^2(S)$ consisting of eigenfunctions of $T$.\\ With $\mathbf{\theta}=(\theta_0,\theta_1,...,\theta_{n-1}) \in S$, we have
\begin{equation}
\begin{aligned}\phi_{0}(\theta) & =\frac{1}{|S|} &  & \text{is an eigenfunction to the eigenvalue} &  & \nu_{0}=1,\\
\phi_{1}^j(\theta) & =\frac{n\theta_j}{|S|} &  & \text{are eigenfunctions to the eigenvalue} &  & \nu_{1}=\int_{S}\ell(\cdot,|\eta-1|)\eta_{1}d\eta<1. \label{eq: eigen}
\end{aligned}
\end{equation}
Any function $\bar{\sigma}\in L^2(\mathds{R}^n\times \mathds{R}^+\times S)$ admits a unique decomposition 
\begin{equation}
\bar{\sigma}=\frac{1}{|S|}\left(\bar{u}+n\mathbf{\theta}\cdot \bar{w} \right)+\hat{z},\label{eq: real_eigen}
\end{equation}
where $\hat{z}$ is orthogonal to all linear polynomials in $\theta$. Explicitly,
\begin{equation*}
\bar{u}(\mathbf{x},t)=\int_S\bar{\sigma}(\mathbf{x},t,\mathbf{\theta})\phi_0(\theta) d\theta,\ \bar{w}^j(\mathbf{x},t)=\int_S \bar{\sigma}(\mathbf{x},t,\mathbf{\theta})\phi_1^j(\theta) d\theta,
\end{equation*}
and  $\bar{w} =  (\bar{w}^1, \dots, \bar{w}^n)$.
\end{lem}

We recall some basic definitions concerning fractional differential operators, as well as their relation to the turning operator $T$.

\begin{defn}\label{def: fractional}
For $s \in (0,2)$ and $f \in C^2(\mathds{R}^n)$ define the fractional gradient of $f$ as
\begin{equation}
\nabla^s f(\mathbf{x})=\frac{1}{|S|}\int_{S}\mathbf{\theta}\mathbf{D}_{\mathbf{\theta}}^s f(\mathbf{x})d\mathbf{\theta}=\frac{1}{|S|}\int_{S}\mathbf{\theta}(\mathbf{\theta}\cdot\nabla)^s f(\mathbf{x})d\mathbf{\theta},\label{eq: fractional derivative}
\end{equation}
where $\mathbf{D}_{\mathbf{\theta}}^s=(\mathbf{\theta}\cdot\nabla)^s$ is the fractional directional derivative of order $s$.
The fractional Laplacian of $f$ is given by
\begin{equation}
\mathds{D}^s f(\mathbf{x})=\frac{1}{|S|}\int_{S}\mathbf{D}^s_\mathbf{\theta}f(\mathbf{x})d\mathbf{\theta}.\label{eq: fractional Laplacian}
\end{equation}
\end{defn}
\noindent It is easily shown that in two dimensions, for $1<\alpha<2$,
\begin{equation}\label{appendixnormal}
\mathds{D}^s = -2\sqrt{\pi} \cos\left(\frac{\pi\alpha}{2} \right)\frac{\Gamma\left(\frac{\alpha+1}{2} \right)}{\Gamma\left(\frac{\alpha+2}{2} \right)}  (-\Delta)^{\nicefrac{s}{2}}\ .
\end{equation}
See Ref.~\refcite{meerschaert2006fractional} for further information.

\bibliographystyle{plain}
\bibliography{Immunology}

\end{document}